\documentclass[twocolumn,prl,showpacs,amsmath,amssymb]{revtex4-1}
\usepackage[dvipdfmx]{graphicx}
\usepackage{bm}
\begin{document}
\title{Logarithmic velocity profile of quantum turbulence of superfluid $^4$He}
\author{Satoshi Yui$^1$}
\author{Kazuya Fujimoto$^1$}
\author{Makoto Tsubota$^{1,2}$}
\affiliation{$^1$Department of Physics, Osaka City University, 3-3-138 Sugimoto, Sumiyoshi-Ku, Osaka 558-8585, Japan}
\affiliation{$^2$The OCU Advanced Research Institute for Natural Science and Technology (OCARINA), Osaka City University, 3-3-138 Sugimoto, Sumiyoshi-Ku, Osaka 558-8585, Japan}
\date{\today}

\begin{abstract}
  The logarithmic velocity profile is the most important statistical law of classical turbulence affected by channel walls. 
  This paper demonstrates numerically that the logarithmic velocity profile of a superfluid flow appears in quantum turbulence under pure normal flow in a channel. 
  We investigated the configuration and dynamics of an inhomogeneous vortex tangle affected by the walls, and found the characteristic behavior of the log-law.
\end{abstract}

\maketitle

The most important statistical laws in the field of classical turbulence are the Kolmogorov -5/3 law of bulk turbulence and the log-law of channel wall \cite{davidson}.
The Kolmogorov -5/3 law of energy spectrum has been confirmed numerically \cite{brachet97,araki,kobayashi} and experimentally \cite{maurer} in quantum turbulence (QT).
This paper is the first report of the log-law in QT.   

QT essentially consists of quantized vortices that are stable topological defects arising from quantum condensation. 
QT is currently one of the most important problems in low temperature physics, and is studied intensively in superfluid helium and atomic Bose--Einstein condensates \cite{halperin,barenghi}.
Almost all studies on QT have been devoted to its bulk behavior.
This study numerically investigated the boundary effect of QT and confirmed the log-law, namely the logarithmic velocity profile, which is established in the field of classical turbulence \cite{marusic}.

A traditional system of QT is thermal counterflow in superfluid $^4$He \cite{donnelly, tough}.
Thermal counterflow is characteristic of the two-fluid model that describes the system as a mixture of viscous normal fluid and inviscid superfluid. 
The relative motion of two fluids is driven by heat injection in a channel. 
When their relative velocity exceeds a critical value, a tangle of quantized vortices appears and grows to make the superfluid flow turbulent.
This scenario was confirmed experimentally \cite{vinen57,tough} and numerically \cite{schwarz88,adachi}.
However, most numerical studies assumed that the flow profile of the normal fluid was uniformly laminar because no useful information was available on what actually happens to the normal fluid in the channel. 

Recent visualization experiments have changed the situation by observing an unexpected inhomogeneous profile of thermal counterflow.
By using a laser-induced fluorescence technique with metastable He$^*_2$ molecules, Guo {\it et al.} \cite{guo10} and Marakov {\it et al.} \cite{marakov} observed an inhomogeneous velocity profile of a normal fluid component in a square channel.
Motivated by this experiment, some researchers numerically studied the inhomogeneous vortex tangle between two parallel plates under the prescribed Hagen--Poiseuille flow \cite{galantucci, baggaley13, baggaley14, khomenko}, and in a square channel under the prescribed Hagen--Poiseuille and tail-flattened normal fluid flow \cite{yui15}.
Vortices near the walls were denser than those in the central region, forming a superfluid boundary layer.  
Yui and Tsubota calculated the flow profile of a superfluid due to the inhomogeneous vortex tangle.

This result reminds us of the logarithmic velocity profile of turbulence in a classical viscous fluid. 
Here, it would be useful to briefly describe the well-known log-law in classical fluid dynamics \cite{davidson}.
Let us consider a turbulent flow between two parallel plates.
We take the direction of the flow as the $x$-axis and the plane of the wall as the $xz$-plane so that $y$ is the distance from one wall. 
The log-law states that the mean value $u$ of the $x$ component of the turbulent velocity obeys 
\begin{equation} \label{log}
  u=\frac{v^*}{\kappa}(\log y + {\rm const.}),
\end{equation}
where $\kappa \sim 0.4$ is known as the Karman constant and $v^{*}$ is a characteristic velocity.
This log-law is obtained from a physical picture of the boundary effect when the viscosity is negligible.
The fluid momentum is transferred from the central part of the channel toward the wall as a constant flux per unit time. 
The transferred momentum is dissipated by the fluid viscosity near the wall. 
The log-law has been confirmed in classical turbulence in a channel \cite{marusic}.

In this paper, we show that the mean velocity of a turbulent superfluid in a channel obeys the log-law. 
In contrast to classical cases, the statistical behavior should be reduced to the configuration and dynamics of quantized vortices.  
To elucidate the effect of the channel walls more clearly than our previous study \cite{yui15}, we considered QT in a pure normal flow between two parallel plates \cite{tough}.

In a vortex filament model, a quantized vortex is represented by a filament passing through a fluid and has a definite vorticity \cite{schwarz85}.
This approximation is very suitable for He II because the core size of a quantized vortex is much smaller than any other characteristic length scale.
At the zero temperature, the vortex filament moves with the superfluid velocity
\begin{equation} \label{super}
  {\bm v}_{\rm s} =
      {\bm v}_{{\rm s},\omega}
    + {\bm v}_{{\rm s},{\rm b}}
    + {\bm v}_{{\rm s},{\rm a}}
    ,
\end{equation}
where ${\bm v}_{{\rm s},\omega}$ is the velocity field produced by vortex filaments, ${\bm v}_{{\rm s},{\rm b}}$ is that produced by solid boundaries, and ${\bm v}_{{\rm s},{\rm a}}$ is the applied flow of the superfluid.
The filament is represented in parametric form as ${\bm s}={\bm s}(\xi,t)$, where $\xi$ is the arc length along the filament.
The velocity field ${\bm v}_{{\rm s},\omega}$ is given by the Biot--Savart law:
\begin{equation}
  {\bm v} _{{\rm s},\omega} ({\bm r})=
    \frac{\gamma}{4 \pi} \int _{\cal L}
      \frac{ ({\bm s} _{1} - {\bm r}) \times d {\bm s} _{1} }{ |{\bm s} _{1} - {\bm r} |^{3} }
      ,
\end{equation}
where $\gamma = 1.00 \times 10^{-7} ~\mathrm{m^2/s}$ is the quantum of circulation and the integration is performed along the filament.
This paper addresses the full Biot--Savart integral \cite{adachi}.
The velocity field ${\bm v}_{{\rm s},{\rm b}}$ is obtained by a simple procedure;
it is the field produced by an image vortex that is constructed by reflecting the filament onto the surface and reversing its direction.
The dynamics at finite temperatures includes the mutual friction; 
the velocity of a point $\bm s$ on the filament is given by
\begin{equation}
  {\dot{\bm s}} = {{\bm v}_{\rm s}}
    + \alpha {\bm s}' \times ({\bm v}_{\rm n} - {{\bm v}}_{\rm s})
    - \alpha '{\bm s}' \times [{\bm s}'\times({\bm v}_{\rm n} - {{\bm v}}_{\rm s})],
  \label{vortex}
\end{equation}
where $\alpha$ and $\alpha'$ are the temperature-dependent coefficients and ${\bm s}'=d{\bm s}/d\xi$ is a unit vector along the filament.

\begin{figure}[htb]
  \includegraphics[width=0.48\textwidth]{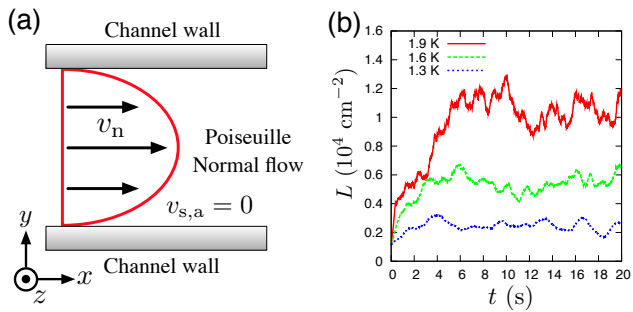}
  \caption
  {
    (Color online)
    (a) Schematics of simulation of pure normal flow.
        The normal fluid component flows along the $x$-axis with a Poiseuille profile between two parallel plates, whereas the superfluid component has no external flow.
        The periodic boundary condition is applied to the $x$ and $z$ directions and solid boundary condition to the $y$ direction.
    (b) Vortex line density as a function of $t$ under pure normal flow with $\overline{v}_{\rm n} = 0.9 ~\mathrm{cm/s}$.
        QT develops to a statistically steady state.
  }
  \label{setup.eps}
\end{figure}

In this paper, we address pure normal flow \cite{tough}.
As shown in Fig. \ref{setup.eps}(a), the normal fluid component flows between two parallel plates, whereas the superfluid component has no external flow, {\it i.e.}, $v_{{\rm s},{\rm a}}=0$.
The flow direction is along the $x$-axis.
The solid boundaries are applied at $y/D= 0$ and $2$ with the half-width $D$ of the channel.
The normal flow is prescribed to be a Poiseuille profile:
\begin{equation}
  \bm{v}_{\rm n} = u_0 \left[ 1- \left( \frac{y-D}{D} \right)^2 \right] \hat{\bm x},
\end{equation}
where $u_0$ is a normalization factor and $\hat{\bm x}$ is a unit vector in the $x$ direction.

Simulations were performed under the following conditions.
We discretized the vortex lines into a number of points held at a minimum space resolution of $\Delta \xi = 8.0 \times 10^{-4} ~\mathrm{cm}$.
Integration in time was achieved using a fourth-order Runge--Kutta scheme with  time resolution $\Delta t = 1.0 \times 10^{-4} ~\mathrm{s}$.
The computing box was $0.1 \times 0.1 \times 0.1 \mathrm{~cm^3}$.
Periodic boundary condition was applied in the $x$ and $z$ directions, whereas solid boundary condition was applied at the walls.
We reconnected two vortices artificially when the vortices approached each other more closely than $\Delta \xi$.
We eliminated vortices that were shorter than $5 \times \Delta \xi = 2.4 \times 10^{-3} ~\mathrm{cm}$.
The temperatures were $T=1.9 ~\mathrm{K} ~(\alpha=0.21, \alpha'=0.009)$, $1.6 ~\mathrm{K} ~(\alpha=0.098, \alpha'=0.016)$, and $1.3 ~\mathrm{K} ~(\alpha=0.036, \alpha'=0.014)$ \cite{schwarz85}.
The initial state consisted of eight randomly oriented vortex rings of radius $0.023 \mathrm{~cm}$.
The simulation was finalized at $t=2.0 \times 10 ~\mathrm{s}$ for $T=1.9 ~\mathrm{K}$ and $1.6 ~\mathrm{K}$, and $t=6.0 \times 10 ~\mathrm{s}$ for $T=1.3 ~\mathrm{K}$.

\begin{figure}[htb]
  \includegraphics[width=0.45\textwidth]{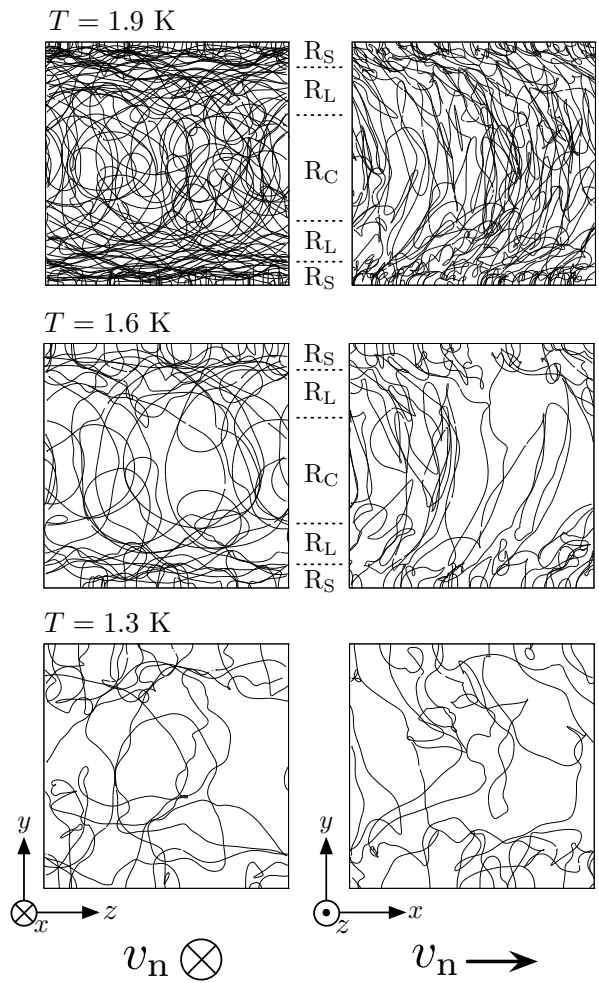}
  \caption
  {
    Snapshots of the vortex tangles in the statistically steady states.
    The left (right) column shows the streamwise (side) views of the tangles.
  }
  \label{tangle.eps}
\end{figure}

The quantized vortex tangle in the pure normal flow develops to a statistically steady state.
Figure \ref{setup.eps}(b) shows the time evolution of the vortex line density $L=\int _{\mathcal L} d{\xi}/\Omega$, where $\Omega$ is the whole volume.
At all temperatures, we applied the normal fluid flow with $\overline{v}_{\rm n} =0.9 ~\mathrm{cm/s}$ to the system, where $\overline{v}_{\rm n}$ is the value of $v_{\rm n}$ averaged over the channel cross section.
The value of $L$ develops from the initial state and then fluctuates around a constant value.
This means that QT is in a statistically steady state.
The statistically steady value of $L$ increases with $T$ because of a stronger mutual friction.
An inhomogeneous vortex tangle appears in this statistically steady state.
Figure \ref{tangle.eps} shows the configurations of the quantized vortices in the statistically steady state \cite{movie}.
The vortices concentrate near the channel walls.
A similar structure was already reported in the simulations of a nonuniform counterflow \cite{baggaley13, baggaley14, yui15, khomenko}.
With increasing $T$, the inhomogeneous structure becomes more apparent and anisotropic because of the stronger mutual friction.
The configuration and dynamics of the tangle were studied by focusing on coarse-grained physical quantities in the latter part of this paper.

\begin{figure}[htb]
  \includegraphics[width=0.48\textwidth]{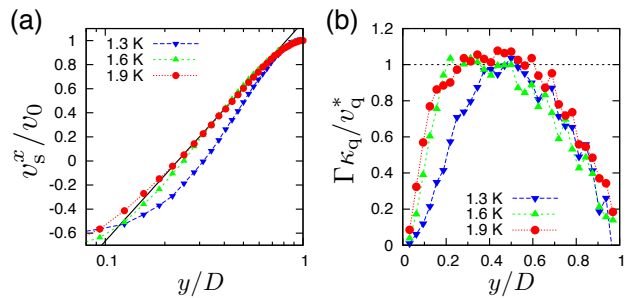}
  \caption
  {
    (Color online)
    (a)
    Flow direction component $v_{\rm s}^x$ of the superfluid velocity ${\bm v}_{\rm s}$ as a function of the distance $y/D$ from the wall.
    The channel wall corresponds to $y/D = 0$, whereas the center of the channel corresponds to $y/D=1$. 
    The log-law $v_{\rm s}^{y} = v_{\rm q}^{*} \left[ \log(y/D) + c \right] /\kappa_{\rm q}$ is observed.
    The solid line is fitting line for $T=1.9 ~\mathrm{K}$.
    (b)
    Another form $\Gamma \kappa_{\rm q}/v_{\rm q}^{*} =1$ of the log-law in Eq. (\ref{log_q_2}) is checked.
    We use the values of $v_{\rm q}^{*}/\kappa_{\rm q}$ in Table \ref{table1}.
  }
  \label{velocity.eps}
\end{figure}

The log-law of the superfluid flow is found in the statistically steady states.
The flow direction component $v_{\rm s}^x/v_0$ of Eq.(\ref{super}) is plotted in Fig. \ref{velocity.eps}(a), where $v_0$ is the value of $v_{\rm s}^x$ at the center $y/D=1$ of the channel.
The values are averaged in the $x$ and $z$ directions and over the statistically steady states.
The data are plotted as a function of the distance $y/D$ from the wall.
We can observe the logarithmic velocity profile:
\begin{equation} \label{log_q_1}
  v_{\rm s}^x=\frac{v_{\rm q}^*}{\kappa_{\rm q}} \left[ \log \left( \frac{y}{D} \right) + c \right],
\end{equation}
where $v_{\rm q}^*$ is a characteristic velocity of QT, $\kappa_{\rm q}$ is a Karman constant for QT, and $c$ is a constant value.
By fitting line to the data, we obtained the values of $v_{\rm q}^* /\kappa_{\rm q}$ and $c$ in Table \ref{table1}.
The quantum Karman constant $\kappa_{\rm q}$ cannot be determined.
To obtain $\kappa_{\rm q}$, we should know the physical meaning of $v_{\rm q}^{*}$, but we have not constructed a theory of the log-law in QT giving $v_{\rm q}^{*}$.

\begin{table}[htb]
  \caption
  {
    Numerical results of superfluid velocity $v_0$ at the center of the channel, ratio $v_{\rm q}^{*} / \kappa _{\rm q}$, and parameter $c$ under pure normal flow.
  }
  \begin{tabular}{cccc}
    \hline\hline
    $T$ & $v_0$  & $v_{\rm q}^{*} / \kappa _{\rm q}$ & $ c $ \\
    (K) & (s/cm) & (s/cm) & -- \\
    \hline
    1.9 & 0.184    & 0.141    & 1.46    \\
    1.6 & 0.079    & 0.070    & 1.40    \\
    1.3 & 0.025    & 0.028    & 1.14    \\
    \hline\hline
  \end{tabular}
  \label{table1}
\end{table}
 

Verifying exactly whether the log-law appears may be difficult.
In fact, researchers of classical turbulence have made many discussions of the log-law; for example, some researchers argue that the mean velocity obeys a power-law \cite{barenblatt, george}.
To analyze the log-law more intensively, we check for another form of the log-law:
\begin{equation} \label{log_q_2}
  y\frac{dv_{\rm s}^{x}}{dy} = \frac{v_{\rm q}^{*}}{\kappa_{\rm q}},
\end{equation}
which is obtained by differentiating Eq. (\ref{log_q_1}) with respect to $y$.
Figure \ref{velocity.eps}(b) plots $\Gamma \kappa_{\rm q} / v_{\rm q}^{*}$ as a function of $y/D$, where $\Gamma \equiv y(dv_{\rm s}^{x} / dy)$ is the left-hand side of Eq. (\ref{log_q_2}).
The $y$-independent regions correspond to the log-law region and can be observed in the case of $T=1.9 ~\mathrm{K}$ and $1.6 ~\mathrm{K}$.
The log-law regions are around $0.2 < y/D < 0.6$.
The log-law region is not wide; therefore, we plan to perform another simulation in the future to expand the log-law region.
As for $T=1.3 ~\mathrm{K}$, the $y$-independent region is too narrow to definitively find the log-law.
Considering the superfluid velocity and the vortex configuration, we can classify the whole volume into three regions, namely the log-law region ${\rm R}_{\rm L}$, the high curvature vortex region ${\rm R}_{\rm S}$ near the wall, and the low vortex line density region ${\rm R}_{\rm C}$ in the center of the channel, as shown in Fig. \ref{tangle.eps}.

The configuration and dynamics of the vortex tangle are important for understanding the log-law in QT because the superfluid velocity is determined only by the quantized vortices.
In the following, we report on our investigation of the configuration and dynamics of the vortex tangle characteristic of the log-law.
We define the coarse-grained value of a physical quantity $Q(\xi)$ as
\begin{equation} \label{inhomo}
   \left[ Q \right]_{\rm CG} ({\bm r})
    \equiv \frac{1}{\omega({\bm r}) l({\bm r})}\int_{\mathcal L'({\bm r})} Q(\xi) d\xi,
\end{equation}
where $\omega({\bm r})$ is the local subvolume at $\bm r$, $\mathcal L'({\bm r})$ represents the vortex line within $\omega({\bm r})$, and $l({\bm r})$ is the local vortex line density at $\bm r$.
We divide the computational box into the subvolumes by using a uniform $32 \times 32 \times 32$ Cartesian mesh.
The separation of the mesh is $\Delta x=1/32 ~\mathrm{mm} = 3.125 \times 10^{-3} ~\mathrm{cm}$; the mesh satisfies $\Delta x > \Delta \xi$.
In the following, to focus on the statistical value, physical quantities are always averaged in the $x$ and $z$ directions and over the statistically steady states.

\begin{figure}[htb]
  \includegraphics[width=0.48\textwidth]{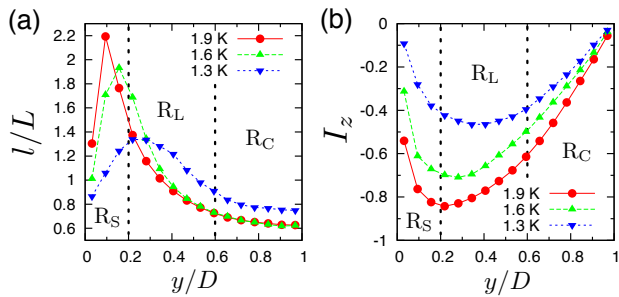}
  \caption
  {
    (Color online)
    Local vortex line density $l$ and anisotropic parameter $I_z$ as a function of the distance $y/D$ from the wall.
  }
  \label{inhomogeneity.eps}
\end{figure}

First, the configuration of the tangle was considered.
The local vortex line density $l$ is plotted in Fig. \ref{inhomogeneity.eps}(a).
The value of $l$ in ${\rm R}_{\rm C}$ is smaller than those in ${\rm R}_{\rm S}$ and ${\rm R}_{\rm L}$, because in ${\rm R}_{\rm C}$ the stronger relative velocity between the two fluids makes the mutual friction terms in Eq. (\ref{vortex}) dominant so that the vortices expand rapidly and escape from ${\rm R}_{\rm C}$.
When approaching the wall from the center, the value of $l$ once increases and then decreases near the wall.
Comparing the data at $T= 1.9 ~\mathrm{K}$ and $1.6 ~\mathrm{K}$ with those at $T=1.3 ~\mathrm{K}$, we find that the characteristic behavior of the log-law may be the strong inhomogeneity of $l$.
On the basis of this observation, the ambiguity of the log-law at $T=1.3 ~\mathrm{K}$ may be attributable to the weak inhomogeneity.
Figure \ref{inhomogeneity.eps}(b) shows the anisotropic parameter $I_i \equiv \left[ {\bm s}'\cdot{\hat{\bm r}_i} \right]_{\rm CG}$, where $\hat{\bm r}_i$ is a unit vector in the $i$ direction.
A large value of $|I_i|$ means that the vortices tend to be parallel to the $i$ direction.
When approaching the wall, the value of $|I_z|$ increases in ${\rm R}_{\rm C}$ and ${\rm R}_{\rm L}$ and decreases in ${\rm R}_{\rm S}$.
Although a clear difference is not observed between ${\rm R}_{\rm C}$ and ${\rm R}_{\rm L}$, we can infer that the log-law needs a larger anisotropy than other regions.

\begin{figure}[htb]
  \includegraphics[width=0.48\textwidth]{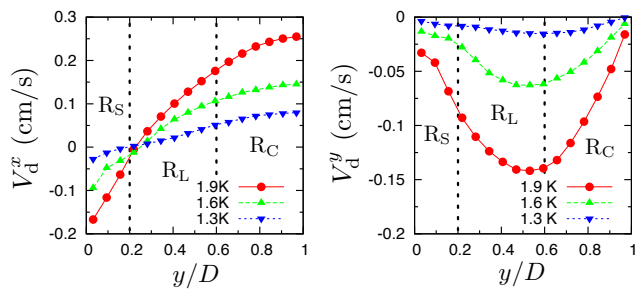}
  \caption
  {
    (Color online)
    Drift velocity $V_d^x$ and $V_d^y$ as a function of the distance $y/D$ from the wall.
  }
  \label{drift.eps}
\end{figure}

Second, we investigated the dynamics of the tangle \cite{movie}.
The drift velocity $V_d^i \equiv \left[ \dot{\bm s}\cdot{\hat{\bm r}}_i \right]_{\rm CG}$ of the vortices is plotted in Fig. \ref{drift.eps}.
The data of $V_{\rm d}^x$ show that in ${\rm R}_{\rm C}$ and ${\rm R}_{\rm L}$ the vortices move in the flow direction of the normal fluid component, whereas in ${\rm R}_{\rm S}$ the vortices move opposite to this direction.
In ${\rm R}_{\rm C}$ and ${\rm R}_{\rm L}$, the superfluid velocity ${\bm v}_{\rm s}$ is along the normal fluid flow as shown in Fig. \ref{velocity.eps}, and in this system, the term including $\alpha '$ in Eq. (\ref{vortex}) tends to carry the vortices in this direction.
In ${\rm R}_{\rm S}$, the value of $|{\bm v}_{\rm n}-{\bm v}_{\rm s}|$ becomes small, so that the mutual friction terms in Eq. (\ref{vortex}) become less dominant and these vortices move against the normal fluid flow by their self-induced velocity.
A clear difference in $V_d^x$ cannot be observed between ${\rm R}_{\rm L}$ and ${\rm R}_{\rm C}$, but between ${\rm R}_{\rm S}$ and ${\rm R}_{\rm L}$ there is difference in the direction of $V_d^x$.
The data of $V_{\rm d}^y$ show that the vortices are carried from ${\rm R}_{\rm C}$ to ${\rm R}_{\rm S}$ by the mutual friction. 
This result and the strong anisotropy $I_z$ in ${\rm R}_{\rm L}$ can cause the momentum transfer of the superfluid component to the wall by the vortices.
According to the analogy with the theory of the log-law in classical turbulence, this mechanism may sustain the log-law in QT.
When approaching the wall, in ${\rm R}_{\rm C}$ the value of $|V_{\rm d}^y|$ increases, and in ${\rm R}_{\rm L}$ and ${\rm R}_{\rm S}$ the value decreases.
Contrary to $V_{\rm d}^x$, a clear difference in the first derivative of $V_{\rm d}^y$ is observed between ${\rm R}_{\rm L}$ and ${\rm R}_{\rm C}$.
These results imply that both $V_{\rm d}^x$ and $V_{\rm d}^y$ are important in the log-law region.

In summary, we demonstrated numerically the logarithmic velocity profile of superfluid flow in QT under pure normal flow in a channel.
To understand the log-law from the configuration and dynamics of the vortex tangle, we investigated the coarse-grained quantities of the vortex line density, anisotropy, and drift velocity.
In the future, we will construct a theory of the log-law in QT and perform another simulation to obtain a wider log-law region.

We would like to acknowledge T. Goto and Y. Tsuji for useful discussions.
M. T. was supported by JSPS KAKENHI Grant No. 26400366 and MEXT KAKENHI ``Fluctuation \& Structure" Grant No. 26103526. 
K.F. was supported by a Grant-in-Aid for 630 JSPS Fellows (Grant No. 262524).


\end{document}